\documentclass[aps,preprint,nofootinbib,hyperref]{revtex4}%
\usepackage{amsfonts}
\usepackage{amsmath}
\usepackage{amssymb}
\usepackage{graphicx}%
\providecommand{\U}[1]{\protect\rule{.1in}{.1in}}
\providecommand{\U}[1]{\protect\rule{.1in}{.1in}}
\providecommand{\U}[1]{\protect\rule{.1in}{.1in}}
\providecommand{\U}[1]{\protect\rule{.1in}{.1in}}
\providecommand{\U}[1]{\protect\rule{.1in}{.1in}}
\providecommand{\U}[1]{\protect\rule{.1in}{.1in}}

\begin{document}
\preprint{ }
\title{Dynamical String Tension in String Theory \\with Spacetime Weyl Invariance\\\bigskip}
\author{Itzhak Bars}
\affiliation{Department of Physics and Astronomy, University of Southern California, Los
Angeles, CA 90089-0484 USA }
\author{Paul J. Steinhardt}
\affiliation{Department of Physics and Princeton Center for Theoretical Physics, Princeton
University, Princeton, NJ 08544, USA}
\author{Neil Turok}
\affiliation{Perimeter Institute for Theoretical Physics, Waterloo, ON N2L 2Y5,
Canada\bigskip}
\affiliation{}

\begin{abstract}
The fundamental string length, which is an essential part of string theory,
explicitly breaks scale invariance. However, in field theory we demonstrated
recently that the gravitational constant, which is directly related to the
string length, can be promoted to a dynamical field if the standard model
coupled to gravity (SM+GR)\ is lifted to a locally scale (Weyl) invariant
theory. The higher gauge symmetry reveals previously unknown field patches
whose inclusion turn the classically conformally invariant SM+GR into a
geodesically complete theory with new cosmological and possibly further
physical consequences. In this paper this concept is extended to string theory
by showing how it can be \textquotedblleft Weyl lifted\textquotedblright\ with
a local scale symmetry acting on target space background fields. In this
process the string tension (fundamental string length) is promoted to a
dynamical field, in agreement with the parallel developments in field theory.
We then propose a string theory in a geodesically complete cosmological
stringy background which suggests previously unimagined directions in the
stringy exploration of the very early universe.

\end{abstract}

\maketitle
\tableofcontents

\newpage

\section{Introduction}

The well known theory of strings propagating in flat or non-trivial
backgrounds is not invariant under local scale (Weyl) transformations in
\textit{target space} since this formulation of strings contains a
dimensionful parameter, namely the string length (equivalently the string
tension or slope parameter $\alpha^{\prime}$), which is closely related to the
gravitational constant in Einstein's general relativity. Our goal in this
paper is to reformulate string theory with a local scale symmetry in target
space without any fundamental lengths so that the string tension $T=\left(
2\pi\alpha^{\prime}\right)  ^{-1}$ emerges from gauge fixing a field to a
constant value.

The motivation for seeking such a formalism is our recent work
\cite{BST-ConfCosm} on the Weyl invariant formulation of the Standard Model
coupled to gravity which led to a classical cosmology \cite{BST-ConfCosm}%
-\cite{BST-sailing} that is geodesically complete across big bang and big
crunch singularities \cite{BST-sailing}. The benefit of introducing the higher
gauge symmetry is the inclusion of missing patches of field space and
spacetime that are not evident in the conventional formalism. The geodesically
complete field space was helpful in discovering new cosmological phenomena
especially in the vicinity of the big bang or big crunch type singularities
\cite{BCST1-antigravity}. To better understand the new phenomena further
analysis is needed in the context of quantum gravity for which string theory
is the leading candidate. Having learned some new lessons in field theory, we
are strongly motivated to study the role of Weyl symmetry in \textit{target
space} in the context of string theory.

The Weyl invariant formulation of the Standard Model coupled to gravity
required the replacement of the gravitational constant by a field with certain
special properties. We should expect a parallel treatment of string theory
where the constant string tension is replaced by a dynamical field. This is
the route we will follow in this paper to introduce the formulation of Weyl
invariance in target space in string theory.

One possible application of such a string theory is to investigate strings
propagating in geodesically complete cosmological backgrounds that include big
crunch and big bang singularities, thus extending our earlier work on
cosmology in the context of field theory. We believe that learning about
string theory and quantum gravity in such backgrounds is essential for
understanding the physics of the very early universe.

\section{Weyl Invariant Low Energy Effective String Theory}

Strings propagating in background fields such as the gravitational metric
$G_{\mu\nu}\left(  X\right)  $, antisymmetric field $B_{\mu\nu}\left(
X\right)  $, dilaton $\Phi\left(  X\right)  $, etc., are described
conventionally with an action that is conformally invariant on the world sheet
with the following form up to order $\alpha^{\prime}$ (see Eqs.(3.4.45-3.4.54)
in \cite{GSW})
\begin{equation}
S_{string}=-\frac{1}{4\pi\alpha^{\prime}}\int d^{2}\sigma\left[
\begin{array}
[c]{c}%
\left(  \sqrt{-h}h^{ab}G_{\mu\nu}\left(  X\right)  +\varepsilon^{ab}B_{\mu\nu
}\left(  X\right)  \right)  \partial_{a}X^{\mu}\partial_{b}X^{\nu}\\
-\alpha^{\prime}\sqrt{-h}R^{\left(  2\right)  }\left(  h\right)  \Phi\left(
X\right)  +\cdots
\end{array}
\right]  \label{Sstring}%
\end{equation}
where the string represented by $X^{\mu}\left(  \sigma^{a}\right)  $ is the
map from the worldsheet to target space in $d$ dimensions, $h_{ab}\left(
\sigma^{a}\right)  $ is the worldsheet metric and $R^{\left(  2\right)
}\left(  h\right)  $ is its Riemann curvature. The dots $\cdots$ correspond to
higher order corrections in powers of the string slope $\alpha^{\prime}$ (to
insure the worldsheet conformal symmetry at the quantum level) as well as to
additional background fields beyond $\left(  G_{\mu\nu},B_{\mu\nu}%
,\Phi\right)  $. The \textquotedblleft slope\textquotedblright\ $\alpha
^{\prime},$ which is inversely proportional to the string tension $T=\left(
2\pi\alpha^{\prime}\right)  ^{-1}$, has the dimensions of $\left(
length\right)  ^{2}.$

The conformal symmetry on the worldsheet requires at the quantum level that
the beta functions corresponding to the couplings $\left(  G_{\mu\nu}%
,B_{\mu\nu},\Phi,\cdots\right)  $ must vanish. The vanishing of the beta
functions correspond to equations of motion for $\left(  G_{\mu\nu},B_{\mu\nu
},\Phi,\cdots\right)  $ that can be derived from the following effective low
energy field theory action in $d$ dimensions,\footnote{The expression given is
for the bosonic string. For superstrings or heterotic strings, the final term
is different, but still zero in the critical dimension~\cite{myers}.}
\begin{equation}
S_{eff}=\frac{1}{2\kappa_{d}^{2}}\int d^{d}x\sqrt{-G}e^{-2\Phi}\left\{
R\left(  G\right)  +4\left(  \partial\Phi\right)  ^{2}-\frac{1}{12}H^{2}%
-\frac{d-26}{3\alpha^{\prime}}+\cdots\right\}  ,\label{Seff}%
\end{equation}
where the completely antisymmetric tensor $H_{\mu\nu\lambda}$ is the curl of
$B_{\mu\nu}$,
\begin{equation}
H_{\mu\nu\lambda}=\partial_{\lbrack\lambda}B_{\mu\nu]}.\label{Hs}%
\end{equation}
In this action the dimensionful gravitational constant parameter $\kappa
_{d}^{2}$, which has dimension $\left(  length\right)  ^{d-2}$, is
proportional to a power of the string slope parameter $\left(  \alpha^{\prime
}\right)  ^{\left(  d-2\right)  /2}.$ The proportionality constant can be
absorbed into a redefinition of the dimensionless string coupling constant
$g_{s}=e^{a}$ which emerges from a constant shift of the dimensionless dilaton
field $\Phi\left(  X\right)  \rightarrow\Phi\left(  X\right)  +a$ as seen in
(\ref{Seff}). By adopting an appropriate definition of a shifted $\Phi$, we
can take
\begin{equation}
\kappa_{d}^{2}=\left(  2\alpha^{\prime}\right)  ^{\left(  d-2\right)
/2}.\label{kappa}%
\end{equation}
This choice is convenient since string computations are often done by using
units in which $2\alpha^{\prime}=1$ which then sets units with $\kappa_{d}=1$
in every dimension.

We now lift the effective string theory action to a Weyl invariant version by
replacing the dimensionful $\kappa_{d}$ (equivalently $\alpha^{\prime}$) by
the expectation value of a field. This step for the effective action will
provide us with the hints for how to lift the string action itself
(\ref{Sstring}) to a Weyl invariant new version of string theory, which is our
ultimate goal in this paper. We are guided by our recent work in
\cite{BST-ConfCosm} of lifting any field theory coupled to gravity to a Weyl
invariant version by replacing the gravitational constant $\kappa_{d}$ with a
field. This process involves an apparent ghost-like scalar field $\phi\left(
x\right)  $ with the wrong sign kinetic energy. Since this field is
compensated by the local scale symmetry (Weyl) there is really no ghost and no
new degree of freedom associated with the extra field. The benefit of
introducing the higher symmetry together with the extra field was discussed in
our papers. Namely, we found that the previous field space is extended to a
geodesically complete spacetime by the inclusion of additional field patches
that were missing since those are not evident in the conventional geodesically
incomplete formulation of gravity in the Einstein frame. Having learned this
lesson in field theory, we extend the idea now first to the effective string
theory $S_{eff}$ and next to the string action $S_{string}$ to formulate a
Weyl invariant version of string theory including background gravitational and
other fields.

We begin with a generalization of the Weyl lifting methods in
\cite{BST-ConfCosm} to $d$ dimensions. Instead of the dilaton $\Phi$ we now
have two scalar fields $\phi^{i}=\left(  \phi,s\right)  $ (one combination is
a gauge degree of freedom) and in addition to the metric $g_{\mu\nu}$ we
include the antisymmetric field $b_{\mu\nu}$ in a Weyl invariant action
$S=\int d^{d}x\mathcal{L}\left(  x\right)  $ given by
\begin{equation}
\mathcal{L}\left(  x\right)  =\sqrt{-g}\left\{
\begin{array}
[c]{c}%
\frac{d-2}{8\left(  d-1\right)  }U\left(  \phi^{k}\right)  \left[  R\left(
g\right)  -\frac{1}{12}H^{2}\left(  b,\phi^{k}\right)  \right] \\
-\frac{1}{2}C_{ij}\left(  \phi^{k}\right)  g^{\mu\nu}\partial_{\mu}\phi
^{i}\partial_{\nu}\phi^{j}-V\left(  \phi^{k}\right)
\end{array}
\right\}  \label{actionWinv}%
\end{equation}
where $H_{\mu\nu\lambda}$ is%
\begin{equation}
H_{\mu\nu\lambda}\left(  b,\phi^{k}\right)  \equiv\partial_{\lbrack\lambda
}b_{\mu\nu]}+ T\left(  \phi^{k}\right)  ^{-1} b_{[\mu\nu}\partial_{\lambda
]}T\left(  \phi^{k}\right)  , \label{HT}%
\end{equation}
where, as we will see, $T\left(  \phi^{k}\right)  $ will play the role of the
dynamical string tension, although at this stage there is no apparent
connection to it. This \textit{modified} $H_{\mu\nu\lambda}$ is constructed to
be invariant, $\delta_{\Lambda}H_{\mu\nu\lambda}=0$ for any $T\left(
\phi\right)  ,$ under the following modified gauge transformation of the
antisymmetric tensor
\begin{equation}
\delta_{\Lambda}b_{\mu\nu}=\partial_{\lbrack\mu}\Lambda_{\nu]}+T\left(
\phi^{k}\right)  ^{-1} \Lambda_{\lbrack\nu}\partial_{\mu]}T\left(  \phi
^{k}\right)  ,
\end{equation}
where the gauge parameter is the vector $\Lambda_{\mu}\left(  x\right)  .$

For \textit{any number of scalars} $\phi^{i},$ requiring this action to be
also invariant under the local scale (Weyl) transformations in $d$ dimensions,%
\begin{equation}
\left(  g_{\mu\nu},b_{\mu\nu}\right)  \rightarrow\Omega^{-\frac{4}{d-2}%
}\left(  g_{\mu\nu},b_{\mu\nu}\right)  ,\text{ and ~}\phi^{i}\rightarrow
\Omega\phi^{i}, \label{WeylTransf}%
\end{equation}
imposes the following homothety conditions on the metric $C_{ij}\left(
\phi\right)  $ in field space \cite{IB-2TSugra}\cite{BST-ConfCosm}%
\begin{equation}
\frac{\partial U}{\partial\phi^{i}}=-2C_{ij}\phi^{i},\;C_{ij}\phi^{i}\phi
^{j}=-U, \label{homothety}%
\end{equation}
and also demands that $\left(  C_{ij},U,V,T\right)  $ be homogeneous functions
of $\phi^{i}$ with homogeneity degrees $\left(  0,2,\frac{2d}{d-2},\frac
{4}{d-2}\right)  $ respectively, namely
\begin{equation}%
\begin{array}
[c]{c}%
C_{ij}\left(  \Omega\phi^{k}\right)  =C_{ij}\left(  \phi^{k}\right)
,\;U\left(  \Omega\phi^{k}\right)  =\Omega^{2}U\left(  \phi^{k}\right)  ,\\
V\left(  \Omega\phi^{k}\right)  =\Omega^{\frac{2d}{d-2}}U\left(  \phi
^{k}\right)  ,\;T\left(  \Omega\phi^{k}\right)  =\Omega^{\frac{4}{d-2}%
}T\left(  \phi^{k}\right)  .
\end{array}
\label{homogeneous}%
\end{equation}
Then $H_{\mu\nu\lambda}$ is covariant under local scale transformations,
$H_{\mu\nu\lambda}\left(  \Omega^{-\frac{4}{d-2}}b,\Omega\phi^{k}\right)
=\Omega^{-\frac{4}{d-2}}H_{\mu\nu\lambda}\left(  b,\phi^{k}\right)  ,$ for any
homogeneous $T$ as indicated above. When $\left(  C_{ij},U,V,T\right)  $
satisfy the conditions (\ref{homothety},\ref{homogeneous}) one can verify that
the Lagrangian (\ref{actionWinv}) transforms into a total derivative under the
local scale transformations, hence the action is invariant.

The general solution of the homothety and homogeneity conditions
(\ref{homothety},\ref{homogeneous}) for any number of scalars $\phi^{i}$, and
in particular for two scalars $\left(  \phi,s\right)  $, is given in
\cite{BST-ConfCosm}. The general solution shows that there remains a lot of
freedom in the choice of the functions $\left(  C_{ij},U,V,T\right)  $ to
construct various models that are invariant under the local scale symmetry.
However, for our case, we will be able to fix all of these functions so that
the string effective action $S_{eff}$ in (\ref{Seff}) emerges when we fix a
gauge for the Weyl gauge symmetry. The gauge of interest here was called the
string gauge or s-gauge in \cite{BCT-cyclic}, where other useful gauges that
will also be useful here were discussed (E-gauge, c-gauge, $\gamma$-gauge).

We now turn to our case of only two scalars $\phi^{i}=\left(  \phi,s\right)
$. By doing field reparametrizations of the $\phi^{i}$ it is always possible
to transform to a basis in which the metric in scalar field space is
conformally flat and of indefinite signature, $ds^{2}=C_{ij}d\phi^{i}d\phi
^{j}$ $\Rightarrow A^{2}\left(  \phi,s\right)  \left(  -d\phi^{2}%
+ds^{2}\right)  .$ In this basis, the homogeneity and homothety conditions on
$U$ and $C_{ij}$ lead to a unique solution for both $U$ and $C_{ij},$ namely
$A^{2}=1$ (constant fixed by normalization) and
\begin{equation}
U=\left(  \phi^{2}-s^{2}\right)  .
\end{equation}
The indefinite metric $C_{ij}$ in field space and the related relative minus
sign in $U$ emerge from physical considerations. It is necessary to allow for
a region or a patch of field space $\left(  \phi,s\right)  $ where the
effective gravitational constant $U\left(  \phi,s\right)  $ is positive in
order to recover ordinary gravity in the Einstein frame when a gauge is fixed.
In the solution for $C_{ij}$ given by $ds^{2}=\left(  -d\phi^{2}%
+ds^{2}\right)  $, the field $\phi$ is a ghost because it has the wrong sign
kinetic energy term in the Lagrangian. If $\phi$ had positive kinetic energy
(like $s$) then $U$ would come out to be purely negative $\left(  -\phi
^{2}-s^{2}\right)  $ which would imply an unacceptable purely negative
gravitational constant in the action (\ref{actionWinv}). This is why there has
to be a relative minus sign and therefore there has to be a ghost in the Weyl
invariant formulation\footnote{There is an additional piece of this argument.
It seems possible to choose a field basis such that $U\left(  \phi,s\right)  $
is always positive, for example, $U\left(  \phi,s\right)  =\left\vert \phi
^{2}-s^{2}\right\vert .$ But then, according to the homothety conditions
(\ref{homothety}), the metric $C_{ij}$ must also have an extra sign,
$ds^{2}=sign\left(  \phi^{2}-s^{2}\right)  \left(  -d\phi^{2}+ds^{2}\right)
.$ More examples of positive $U$ but complicated $C_{ij}\left(  \phi,s\right)
$ are also possible as shown in \cite{BST-ConfCosm}. However, the problem with
such alternative field bases with positive $U$ is that they are all
geodesically incomplete, as demonstrated in our work with explicit analytic
solutions of the field equations. So $U$ must be allowed to change sign.}.
However, the local scale symmetry in the action (\ref{actionWinv}) is just
sufficient to remove this ghost by a gauge choice, so this ghost is not
dangerous for unitarity and therefore it is of no concern.

Hence, for two scalars, using the field basis described above, the Weyl
invariant effective action (\ref{actionWinv}) becomes%
\begin{equation}
\mathcal{L}\left(  x\right)  =\sqrt{-g}\left[
\begin{array}
[c]{c}%
\frac{d-2}{8\left(  d-1\right)  }\left(  \phi^{2}-s^{2}\right)  \left(
R\left(  g\right)  -\frac{1}{12}H^{2}\left(  b,\phi,s\right)  \right) \\
+\frac{1}{2}\partial\phi\cdot\partial\phi-\frac{1}{2}\partial s\cdot\partial
s-V\left(  \phi,s\right)
\end{array}
\right]  \label{actionWinv2}%
\end{equation}
The reader can recognize that in this basis $\left(  \phi,s\right)  $ are both
conformally coupled scalars, thus making the Weyl symmetry more evident. There
remains two unknown functions, $V\left(  \phi,s\right)  $, $T\left(
\phi,s\right)  ,$ that are constrained by homogeneity as in (\ref{homogeneous}).

We now want to show how to recover the low energy effective string action
$S_{eff}$ in Eq.(\ref{Seff}) by choosing the so called \textquotedblleft
string gauge\textquotedblright\ for the local gauge symmetry. In this gauge we
label all fields with an extra label $s$ to indicate that we are in the string
gauge (we will be interested also in other gauge choices labeled by different
symbols, $E,c,\gamma,f$). Thus, in the field patches that satisfy $\left(
\phi^{2}-s^{2}\right)  >0,$ which we call the $\pm$ gravity patches, we
parametrize the fields $\phi,s$ in terms of a single scalar field $\Phi$ (one
degree of freedom is lost in the fixed gauge) and identify the metric and
antisymmetric tensor in the s-gauge with the string backgrounds $G_{\mu\nu
},B_{\mu\nu}$ that appear in $S_{eff}$ and $S_{string}$%
\begin{equation}%
\begin{array}
[c]{c}%
g_{\mu\nu}^{s}=G_{\mu\nu},\;b_{\mu\nu}^{s}=B_{\mu\nu},\;\\
\phi_{s}=\pm\sqrt{\frac{4\left(  d-1\right)  }{\left(  d-2\right)  }}%
\frac{e^{-\Phi}}{\kappa_{d}}\cosh\frac{\Phi}{\sqrt{d-1}},\;~\\
s_{s}=\pm\sqrt{\frac{4\left(  d-1\right)  }{\left(  d-2\right)  }}%
\frac{e^{-\Phi}}{\kappa_{d}}\sinh\frac{\Phi}{\sqrt{d-1}}.
\end{array}
\label{s-gauge}%
\end{equation}
where the $\pm$ is needed to cover all regions of field space $\left(
\phi,s\right)  $ in the $\pm$ gravity patches in the $\left(  \phi,s\right)  $
plane as seen in Fig. 1. It is also possible to replace $s_{s}$ by $-s_{s}$ in
Eq.(\ref{s-gauge}) for another parametrization of the gauge choice as we will
discuss below. Although the $\pm$ signs cancel out in the Lagrangian, it does
not cancel out in the equations of motion, so solutions need to be continuous
as $\phi$ or $s$ vanish and change sign. The s-gauge in (\ref{s-gauge})
corresponds to a pair of curves in the $\left(  \phi,s\right)  $ plane where
points on the curves are parametrized by $\Phi$ according to $\left(  \phi
_{s}\left(  \Phi\right)  ,s_{s}\left(  \Phi\right)  \right)  $ as shown in
Fig. 1.%

\begin{figure}
[tbh]
\begin{center}
\includegraphics[
height=2.4344in,
width=2.4344in
]%
{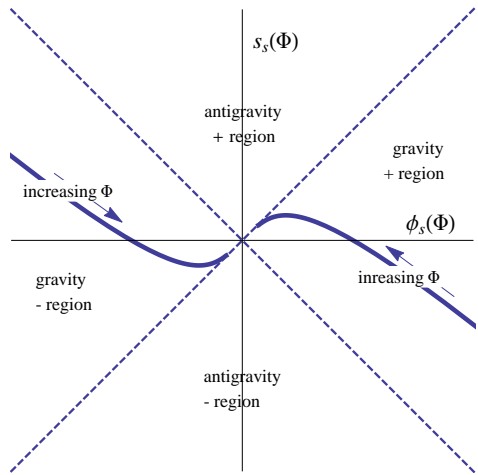}%
\caption{Two branches of the curve in the parametric plot of the string gauge
$\left(  \phi_{s}\left(  \Phi\right)  ,s_{s}\left(  \Phi\right)  \right)  $ in
the $\pm$ gravity regions are shown. The origin corresponds to $\Phi=+\infty$
while the far left/right ends of the curve correspond to $\Phi=-\infty.$ Each
of the four wedge-shaped patches of field space is geodesically incomplete.
The boundary of the gravity/antigravity regions, defined by $\left\vert
s\right\vert =\left\vert \phi\right\vert ,$ is shown with dashed lines.
Geodesically complete solutions of the equations of motion cannot be
constructed in the s-gauge with only the branches of the curve shown in Fig.1.
Geodesic completeness requires additional branches in the antigravity
regions.}%
\label{Fig. 1}%
\end{center}
\end{figure}

If $s_{s}$ is replaced by $-s_{s}$ in (\ref{s-gauge}) we get a new set of
curves in the $\pm$ gravity regions that are similar to the pair of curves in
Fig. 1, but mirror-reflected through the horizontal axis ($s\rightarrow-s$).

Later we will discuss the possibility of including patches in which $\left(
\phi^{2}-s^{2}\right)  $ is negative by simply interchanging $cosh$ and $sinh$
that modifies the gauge choice above leading to curves that look like 90$^{o}$
rotated in relation to those described above. These extend the domain of the
string gauge into patches of antigravity (negative gravitational
\textquotedblleft constant\textquotedblright, as seen from (\ref{actionWinv2}%
)) which are required for geodesically complete cosmological spacetimes as
discussed in a similar setting in \cite{BCT-cyclic}-\cite{BST-sailing}. The
existence of such additional patches is not evident in the effective string
action (\ref{Seff}), but the study of particle geodesics shows that the
spacetime described by (\ref{Seff}) is geodesically incomplete, and this is a
hint that invites a study of the complete space by including antigravity
patches, as discussed below.

With this choice of s-gauge in the gravity patch we see that the coefficients
of the curvature $R$ in (\ref{actionWinv2}) and (\ref{Seff}) match each other,
and furthermore the kinetic energy terms of $\left(  \phi_{s},s_{s}\right)  $
in (\ref{actionWinv2}) precisely reproduce the kinetic energy term of the
dilaton $\Phi$ in (\ref{Seff}), with its unusual normalization, $\sqrt
{-G}\frac{e^{-2\Phi}}{2\kappa_{d}^{2}}4\left(  \partial\Phi\right)  ^{2}$. In
order to have also $H_{\mu\nu\lambda}^{s}$ match the expression in (\ref{Hs}),
it is necessary to determine the homogeneous function $T\left(  \phi,s\right)
$ such that in the string gauge (\ref{s-gauge}) it reduces to a constant as a
function of $\Phi,$ namely $T\left(  \phi_{s}\left(  \Phi\right)
,s_{s}\left(  \Phi\right)  \right)  $=constant with either set of $\pm$ signs
in (\ref{s-gauge}). We find that before gauge fixing, and in the gravity patch
$\left(  \phi^{2}-s^{2}\right)  >0,$ the homogeneous $T\left(  \phi,s\right)
$ of degree $4/\left(  d-2\right)  $ must be given by
\begin{equation}
T\left(  \phi,s\right)  =\left(  \frac{d-2}{4\left(  d-1\right)  }\right)
^{\frac{2}{d-2}}\left(  \phi+s\right)  ^{2\frac{1+\sqrt{d-1}}{d-2}}\left(
\phi-s\right)  ^{2\frac{1-\sqrt{d-1}}{d-2}},\label{T}%
\end{equation}
or the same expression with $s$ replaced by $-s$ in Eq.(\ref{T}) as well as
Eq.(\ref{s-gauge}). This $T\left(  \phi,s\right)  $ is normalized such that it
reduces to the constant string tension in the gravity patches in the s-gauge
given above (recall (\ref{kappa}))
\begin{equation}
\pi T\left(  \phi_{s}\left(  \Phi\right)  ,s_{s}\left(  \Phi\right)  \right)
=\left(  \kappa_{d}^{2}\right)  ^{-\frac{2}{d-2}}=\frac{1}{2\alpha^{\prime}}~.
\end{equation}
As already mentioned, in antigravity patches where $\left(  \phi^{2}%
-s^{2}\right)  <0$ we must exchange $cosh$ and $sinh$ in equation
(\ref{s-gauge}). We may then choose signs in (\ref{s-gauge}) to ensure that
the string tension $T$ as defined in (\ref{T}) behaves smoothly across the
boundary between gravity and antigravity regions. Notice that the overall
constant or sign in front of the tension is irrelevant at this point since
only $T^{-1}\partial_{\lambda}T$ (which is insensitive to the sign of $T$)
appears in $H_{\mu\nu\lambda}$.

It is notable that in the critical dimension $d=10$ for superstrings or
heterotic strings, we may choose signs so that the power of $\phi+s$ (or
$\phi-s$) appearing in the dynamical string tension (\ref{T}) is unity, for
both the gravity (or antigravity) patches. This means that, in a suitable Weyl
gauge, smooth cosmological transitions from gravity to antigravity, or vice
versa, are associated with a smooth analytic sign change of the dynamical
string tension $T$. Such transitions do in fact occur in generic homogeneous
cosmologies \cite{BCST2-solutions}, in a suitable Weyl gauge, termed $\gamma
$-gauge, which will shortly be explained. We emphasize that the signs of
$\left(  \phi+s\right)  $ and $\left(  \phi-s\right)  $ are Weyl gauge
invariants, so their signs in every gauge are the same as those displayed in
the geodesically complete $\gamma$-gauge solution displayed below.

Finally, the potential energy that matches $V\left(  \phi_{s},s_{s}\right)
=\frac{d-26}{3\alpha^{\prime}}\frac{e^{-2\Phi}}{2\kappa_{d}^{2}},$ is also
determined such that before gauge fixing $V\left(  \phi,s\right)  $ is given
by
\begin{equation}
V\left(  \phi,s\right)  =\frac{\left(  d-26\right)  \left(  d-2\right)
}{12\left(  d-1\right)  }\left(  \phi^{2}-s^{2}\right)  T\left(
\phi,s\right)  . \label{V}%
\end{equation}
This $V\left(  \phi,s\right)  $ is homogeneous of degree $\frac{2d}{d-2}$ as
required in (\ref{homogeneous}), it vanishes for $d=26$ and is negative for
$d<26$ in the gravity patch. It is expected that higher perturbative terms in
$\alpha^{\prime}$, non-perturbative,\ and string-string interaction effects
will alter the effective low energy potential. Furthermore in supersymmetric
or heterotic versions of string theory $V$ is not the same as above (e.g.
$\left(  d-26\right)  $ is replaced by $\left(  d-10\right)  $ in SUSY). In
any case it is evident that the final phenomenologically significant result
for $V\left(  \phi,s\right)  $ can always be rewritten in a Weyl covariant
form as, $V=\phi^{\frac{2d}{d-2}}f\left(  s/\phi\right)  ,$ for some Weyl
invariant $f\left(  s/\phi\right)  $ as in our previous work.

By determining the functions $T\left(  \phi,s\right)  $ and $V\left(
\phi,s\right)  $ as in (\ref{T},\ref{V}), we have completely fixed the
effective low energy string action as a Weyl invariant action in the gravity
patch $\left(  \phi^{2}-s^{2}\right)  \geq0$ in the form given in
Eqs.(\ref{actionWinv2},\ref{HT}). A significant property of the new Weyl
invariant effective action is that it becomes geodesically complete by
including all field patches in the infinite $\left(  \phi,s\right)  $ plane in
the sense discussed in our previous papers and later in this paper.

To complete the definition of the geodesically complete string action in this
sense we need to weigh carefully the possibilities of whether the smooth
continuation of the functions $T\left(  \phi,s\right)  $ and $V\left(
\phi,s\right)  $ to all patches will change sign when crossing the transition
lines $\left\vert \phi\right\vert =\left\vert s\right\vert $ in the $\left(
\phi,s\right)  $ plane which separate gravity/antigravity patches. This will
be discussed below in the context of the string action with a dynamical
tension $T\left(  \phi,s\right)  $ after a description of geodesic
completeness across gravity/antigravity boundaries given below.

\subsection{Gauge choices for local scale symmetry and geodesic completeness}

The Weyl invariant action (\ref{actionWinv2}) has no dimensionful constants of
any kind. Besides the string gauge given in (\ref{s-gauge}) where the
gravitational constant (or the string tension) is introduced as the gauge
fixed value of a combination of fields, the same action can be gauge fixed to
other gauges that are useful for various physical applications especially in
cosmology as discussed in \cite{BC-inflation}-\cite{BST-sailing}. Some of
these gauges will be useful in string theory as well, so we review here some
convenient gauge choices of the geodesically complete Weyl invariant action
(\ref{actionWinv2}), and then describe the nature of geodesic completeness in
each gauge.

\begin{itemize}
\item \textbf{s-gauge}: This is the \textquotedblleft string
gauge\textquotedblright\ we used above in Eq.(\ref{s-gauge}). A property of
this gauge is that it satisfies $\left\vert \phi_{-s}\right\vert
=c_{d}\left\vert \phi_{+s}\right\vert ^{\lambda_{d}}$ where $c_{d},\lambda
_{d}$ are some constants, and $\phi_{\pm s}\equiv\phi_{s}\pm s_{s}$. More
precisely, as seen from (\ref{s-gauge}), with the choice of
\begin{equation}
c_{d}=\left(  \frac{\kappa_{d}^{2}\left(  d-2\right)  }{4\left(  d-1\right)
}\right)  ^{\frac{1}{\sqrt{d-1}-1}},\;\lambda_{d}=\frac{\sqrt{d-1}+1}%
{\sqrt{d-1}-1},
\end{equation}
$\phi_{s}\left(  \Phi\right)  ,s_{s}\left(  \Phi\right)  $ are parametrized in
the $\pm$ gravity patches in terms of a single field $\Phi$ as in
(\ref{s-gauge}) so that the Weyl invariant action (\ref{actionWinv2}) is
reduced to the gauge fixed action that matches the effective string action
$S_{eff}$ in (\ref{Seff}). Other choices of $\left(  c_{d},\lambda_{d}\right)
$ would yield some similarly gauge fixed action, but a different one than
(\ref{Seff}). The expressions in (\ref{s-gauge}) by themselves are
insufficient to express the gauge choice in all regions of the $\left(
\phi,s\right)  $ plane. These equations must be supplemented by similar
expressions that include more branches of the curve in Fig. 1. The additional
branches correspond to curves obtained from Fig. 1 by a reflection through the
horizontal axis generated by $s\rightarrow-s$ (new curve still in gravity),
plus those obtained by a 90$^{o}$ rotations of the curves already mentioned,
leading to additional branches in the antigravity patches. Later we will
introduce a \textquotedblleft flip symmetry\textquotedblright\ in
Eq.(\ref{flip}) which relates to a symmetry that interchanges gravity and
antigravity. To express the generic geodesically complete cosmological
solution, given in Fig. 2 \ and Eqs.(\ref{metric}-\ref{phi-s}) below in the
$\gamma$-gauge as a function of conformal time $x^{0}$, we need all of those
additional branches in the string gauge basis $\left(  \phi_{s}\left(
\Phi\left(  x^{0}\right)  \right)  ,s_{s}\left(  \Phi\left(  x^{0}\right)
\right)  \right)  $. Accordingly, the geodesically complete form of Fig. 1
must include connections among all the branches of the curves described above,
such that any given branch connects continuously to another branch across the
gravity$/$antigravity intersection by passing through the origin in Fig. 1
tangentially to one of the dashed lines while being perpendicular to the other.

\item \textbf{E-gauge:} This gauge is useful to rewrite the theory in the
Einstein frame where the main physical intuition about gravitational phenomena
is developed. Starting in the $\pm$ gravity patches, it is defined by gauge
fixing $\left(  \phi^{2}-s^{2}\right)  $ to a positive constant,
$\frac{\left(  d-2\right)  }{8\left(  d-1\right)  }\left(  \phi_{E}^{2}%
-s_{E}^{2}\right)  =\frac{1}{2\tilde{\kappa}_{d}^{2}}.$ For the kinetic term
of the remaining scalar (called $\sigma$) to be normalized conventionally,
$-\frac{1}{2}\left(  \partial\sigma\right)  ^{2},$ the fields $\left(
\phi,s\right)  $ are parametrized in terms of the field $\sigma$ as follows
\begin{equation}%
\begin{array}
[c]{c}%
\phi_{E}\left(  \sigma\right)  =\pm\sqrt{\frac{4\left(  d-1\right)  }{\left(
d-2\right)  }}\frac{1}{\tilde{\kappa}_{d}}\cosh\left(  \frac{\left(
d-2\right)  \tilde{\kappa}_{d}}{4\left(  d-1\right)  }\sigma\right)  ,\;~\\
s_{E}\left(  \sigma\right)  =\left(  \pm\text{ or }\mp\right)  \sqrt
{\frac{4\left(  d-1\right)  }{\left(  d-2\right)  }}\frac{1}{\tilde{\kappa
}_{d}}\sinh\left(  \frac{\left(  d-2\right)  \tilde{\kappa}_{d}}{4\left(
d-1\right)  }\sigma\right)  ,
\end{array}
\label{E-gauge}%
\end{equation}
The geometrical fields are also labeled with the letter $E:$ $\left(
g_{\mu\nu}^{E},b_{\mu\nu}^{E}\right)  .$ In this gauge the tension $T\left(
\phi_{E},s_{E}\right)  $ obtained from (\ref{T}) is not a constant and
therefore $H_{\mu\nu\lambda}^{E}$ as defined in (\ref{HT}) has a non-trivial
structure\footnote{A straightforward Weyl rescaling of (\ref{Seff}) to the
Einstein frame in which only $G_{\mu\nu}$ is rescaled is also possible. But
only the transformation of the metric produces a different expression for the
transformed action. This is because the untransformed $H_{\mu\nu\lambda}$ or
$B_{\mu\nu}$ differ form our definitions of $H_{\mu\nu\lambda}^{E}$ and
$b_{\mu\nu}^{E}$ by the extra terms involving $b_{[\mu\nu}^{E}\partial
_{\lambda]}\ln\left\vert T^{E}\right\vert .$ After writing $b_{\mu\nu}^{E}$ in
terms of $B_{\mu\nu}$ and expressing $\sigma$ in terms of $\Phi$ (see below)
we do recover the same results.}. The E-gauge in the gravity patches alone, as
given in (\ref{E-gauge}), is geodesically incomplete because one can find
classical solutions of the fields in which the gauge invariant quantity
$\left(  1-s^{2}/\phi^{2}\right)  $ changes sign as a function of time. This
happens when both $\left(  \phi_{E},s_{E}\right)  $ are infinitely large (or
$\sigma\rightarrow$large) so that $\left(  1-s_{E}^{2}/\phi_{E}^{2}\right)  $
can vanish even though $\left(  \phi_{E}^{2}-s_{E}^{2}\right)  $ is a constant
all the way to the boundary of the gravity/antigravity patches. Hence the
gravity patch by itself is incomplete. This is confirmed by noting that
particle geodesics reach cosmological singularities in a finite amount of
time. Furthermore, solutions of the field equations that extend to the
gravity/antigravity boundary (see Fig. 2) show that when the gauge invariant
$\left(  1-s_{E}^{2}\left(  x\right)  /\phi_{E}^{2}\left(  x\right)  \right)
\rightarrow0$ the spacetime metric $g_{\mu\nu}^{E}\left(  x\right)  $ has also
a curvature singularity in the E-gauge. Hence the transition from gravity to
antigravity patches occurs only at spacetime singularities in the Einstein
frame. To obtain the gauge fixed expressions for $\left(  \phi,s\right)  $ in
the antigravity patches in the E-gauge the \textit{cosh} and \textit{sinh} in
(\ref{E-gauge}) are interchanged; then the kinetic term of $\sigma$ also
changes sign.

\item \textbf{c-gauge:} This gauge is useful to relate the Weyl invariant
theory to low energy degrees of freedom \cite{2Tgravity}, and was used in the
construction and analysis of the standard model coupled to gravity
\cite{BST-ConfCosm}\cite{BST-HiggsCosmo}. It is introduced by gauge fixing the
field $\phi$ to a constant, $\phi_{c}\left(  x\right)  =c,$ and then
$s_{c}\left(  x\right)  $ is the remaining dynamical scalar, while the other
fields are labeled as $\left(  g_{\mu\nu}^{c},b_{\mu\nu}^{c}\right)  $. When
$s_{c}$ is much smaller than the constant $\phi_{c},$ which is true at
energies much smaller than the Planck scale, the factor $\left(  \phi_{c}%
^{2}-s_{c}^{2}\left(  x\right)  \right)  $ is practically a constant from the
perspective of low energy physics. This gauge is geodesically complete because
$\left(  \phi_{c}^{2}-s_{c}^{2}\left(  x\right)  \right)  $ can change sign
dynamically (as seen in explicit solutions, including Fig. 2) when the
spacetime $g_{\mu\nu}^{c}$ transits between gravity and antigravity at
curvature singularities.

\item $\gamma$\textbf{-gauge:} This gauge has been the most useful in
simplifying equations and leading to analytic results. The determinant of the
spacetime metric $g_{\mu\nu}^{\gamma}$ is fixed to be unimodular, $\det\left(
g^{\gamma}\right)  =-1$, while the other fields are labeled with $\gamma$ as
$\left(  \phi_{\gamma},s_{\gamma},b_{\mu\nu}^{\gamma}\right)  .$ For example
for a cosmological spacetime of the form $ds_{\gamma}^{2}=a_{\gamma}%
^{2}\left(  \tau\right)  \left(  -d\tau^{2}+ds_{d-1}^{2}\right)  $ where
$\tau$ is the conformal time, and $ds_{d-1}^{2}$ includes space curvature and
anisotropy, the Weyl symmetry is used to gauge fix the scale factor to a
constant for all conformal time, $a_{\gamma}\left(  \tau\right)  =1.$ The
$\gamma$-gauge is geodesically complete. It was used in cosmological
applications in \cite{BC-inflation}-\cite{IB-BibBang} and led to the discovery
of the geodesically complete nature of spacetime across big bang and big
crunch singularities where the gravity-antigravity transition occurs
\cite{BCST1-antigravity}\cite{BST-sailing}.

\item $f\left(  R\right)  $\textbf{ gravity gauge}: If we choose $\left(
\phi-s\right)  $ to be a constant, $\phi_{f}\left(  x\right)  -s_{f}\left(
x\right)  =c,$ then the kinetic term for the remaining field $\phi_{+}\left(
x\right)  \equiv\phi_{f}\left(  x\right)  +s_{f}\left(  x\right)  $ drops out
of the action (\ref{actionWinv2}). Ignoring the $H^{2}$ term in
(\ref{actionWinv2}), the field $\phi_{+}\left(  x\right)  $ becomes a purely
algebraic field that can be determined via the equations of motion (for any
$V$) to be a function of the curvature $\phi_{+}=\phi_{+}\left(  R\left(
g^{f}\right)  \right)  .$ Inserting this solution back into the action
(\ref{actionWinv2}) reduces it to a function of only $R.$ This is $f\left(
R\right)  $ gravity, where the function $f\left(  R\right)  $ is determined by
$V\left(  \phi,s\right)  $ (see also \cite{bamba}).
\end{itemize}

The transformations of fields from one fixed gauge to another is easily
obtained by considering gauge invariants under the Weyl transformations. Some
useful gauge invariants that may be used for this purpose are
\begin{equation}
\frac{s}{\phi},\;\left(  \sqrt{-g}\right)  ^{\frac{d-2}{2d}}\phi,\;\left(
\sqrt{-g}\right)  ^{\frac{d-2}{2d}}s,\;\left\vert T\left(  \phi,s\right)
\right\vert ^{-\frac{d-2}{4}}\phi,\;\left\vert T\left(  \phi,s\right)
\right\vert ^{-\frac{d-2}{4}}s,\;\text{etc.}%
\end{equation}
To illustrate this, consider the example of the s-gauge versus the E-gauge. By
comparing the gauge invariant, $\frac{s_{s}\left(  \Phi\right)  }{\phi
_{s}\left(  \Phi\right)  }=\frac{s_{E}\left(  \sigma\right)  }{\phi_{E}\left(
\sigma\right)  },$ we obtain the relation between the dilaton $\Phi\left(
x\right)  $ in the s-gauge (\ref{s-gauge}) and the scalar field $\sigma\left(
x\right)  $ in the E-gauge (\ref{E-gauge}). This provides the local scale
transformation $\Omega_{sE}\left(  x\right)  $ that relates the s and E gauges
according to Eq.(\ref{WeylTransf}), and using it we find the relation between
the remaining degrees of freedom, $\left(  G_{\mu\nu},B_{\mu\nu}\right)
=\left(  \Omega_{sE}\right)  ^{-\frac{4}{d-2}}\left(  g_{\mu\nu}^{E},b_{\mu
\nu}^{E}\right)  $. Thus, solutions of equations of motion in one gauge can be
used to obtain solutions in all other gauges via the Weyl transformations
among fixed gauges.

Such field transformations are \textit{duality transformations}: in each gauge
the form of the action, the spacetime and the dynamics appear different, but
the Weyl invariant physical information is identical. Indeed the stringy
T-duality of the effective string action discussed in \cite{Veneziano} is an
automatic outcome of the Weyl invariant formalism discussed here. This is
because the dualities based on inversion of scale factors in different
directions \cite{Veneziano} can be re-stated as a combination of general
coordinate reparametrizations and local Weyl transformations, both of which
are symmetries of our low energy string formalism in (\ref{actionWinv2}). In
our setting here we can construct a richer set of dualities based on Weyl
transformations between different gauge fixed versions of the same Weyl
invariant effective low energy string theory (equivalently, dual backgrounds
for string theory on the worldsheet)\footnote{Weyl symmetry follows from
2T-physics in 4+2 dimensions as a requirement for its conformal shadow in 3+1
dimensions. Indeed, Weyl rescalings in 3+1 are simply local reparametrizations
of the extra 1+1 dimensions as a function of the 3+1 dimensions
\cite{2Tgravity}. In this connection see \cite{AB1-dualities} where an even
larger concept of dualities in phase space (beyond just position space), that
follow from 2T-physics, is discussed. \label{foot2T}}.

Having shown that the low energy effective string theory (\ref{Seff}) is just
a gauge choice of the Weyl-lifted low energy string theory (\ref{actionWinv2}%
), we now are in a position to argue that by allowing the fields to take
values in the full $\left(  \phi,h\right)  $ plane, including the antigravity
regions shown in Fig. 1, we obtain a geodesically complete theory. To see
this, we solve analytically the cosmological equations of motion that are
predicted by (\ref{actionWinv2}) as shown in the next paragraph, and find that
the dynamics requires that solutions that begin initially anywhere in the
gravity patches (such as those shown in Fig. 1) evolve inevitably into the
antigravity patches, thus requiring the inclusion of all patches. Then a
geodesically complete field space includes all of the $\left(  \phi,s\right)
$ plane as well as an extension of the spacetime metric $g_{\mu\nu}\left(
x\right)  $ as a function of spacetime $x^{\mu}$ to include the antigravity
patches. In this complete field space geodesics (appropriately defined to be
Weyl invariant as consistent solutions of the same theory \cite{BST-sailing})
are complete curves that do not artificially end at cosmological singularities
\cite{BST-sailing}. Hence we have argued that the Weyl invariant low energy
string theory (\ref{actionWinv2}) which includes all field patches is
geodesically complete, at least for spacetimes of interest in cosmology.

The behavior of the general solution in the vicinity of cosmological
singularities is unique due to an attractor mechanism discovered in
\cite{BCST1-antigravity}. It can be argued that the generic solution is
kinetic energy dominated so that the potential energy $V$ is negligible.
Taking also $b_{\mu\nu}=0$ for simplicity, but adding conformally invariant
massless fields (treated as radiation $\rho_{r}/a^{4}\left(  x^{0}\right)  $
), we display the solution obtained in \cite{BCST1-antigravity} in four
dimensions, $d=4,$ as follows. For our purposes here it is most simply
described in the $\gamma$-gauge. The metric $g_{\mu\nu}^{\gamma}$ is diagonal
and generically anisotropic near singularities and is parametrized as
\begin{align}
ds^{2}  &  =a_{\gamma}^{2}\left(  -\left(  dx^{0}\right)  ^{2}+\sum_{i=1}%
^{3}e_{i}^{2}\left(  dx^{i}\right)  ^{2}\right)  ,\;a_{\gamma}^{2}=1\text{
(}\gamma\text{-gauge}),\label{metric}\\
e_{1}^{2}  &  =e^{2\left(  \alpha_{1}+\sqrt{3}\alpha_{2}\right)  },e_{2}%
^{2}=e^{2\left(  \alpha_{1}-\sqrt{3}\alpha_{2}\right)  },e_{3}^{2}%
=e^{-4\alpha_{1}},
\end{align}
where $\alpha_{1,2}\left(  x^{0}\right)  $ are two dimensionless geometrical
fields that parametrize the anisotropy such that $e_{1}^{2}e_{2}^{2}e_{3}%
^{2}=1,$ so the metric $g_{\mu\nu}^{\gamma}$ is unimodular. The general
generic solution for $\alpha_{1},\alpha_{2},\phi_{\gamma},s_{\gamma}$ is
\cite{BCST1-antigravity}
\begin{align}
\alpha_{1}  &  =\frac{p_{1}}{2p}\ln\left\vert \frac{x^{0}}{l_{1}^{4}\rho
_{r}(x^{0}-x_{c}^{0})}\right\vert ,\;\label{alpha1}\\
\alpha_{2}  &  =\frac{p_{2}}{2p}\ln\left\vert \frac{x^{0}}{l_{2}^{4}\rho
_{r}(x^{0}-x_{c}^{0})}\right\vert ,\;\label{alpha2}\\
\phi_{\gamma}+s_{\gamma}  &  =l_{3}^{2}\rho_{r}(x^{0}-x_{c}^{0})\left\vert
\frac{x^{0}}{l_{3}^{4}\rho_{r}(x^{0}-x_{c}^{0})}\right\vert ^{(p+p_{3}%
)/2p},\;\label{phi+s}\\
\phi_{\gamma}-s_{\gamma}  &  =\frac{2x^{0}}{l_{3}^{2}}\left\vert \frac{x^{0}%
}{l_{3}^{4}\rho_{r}(x^{0}-x_{c}^{0})}\right\vert ^{-(p+p_{3})/2p}.
\label{phi-s}%
\end{align}
Here $x^{0}$ is conformal time as seen from the definition of the metric in
(\ref{metric}). The constant parameters $\left(  l_{1},l_{2},l_{3}\right)  $
are constants of integration that have dimension of $\left(  length\right)  $;
the parameters $\left(  p_{1},p_{2},p_{3}\right)  $ are constants of
integration that have dimension of $\left(  length\right)  ^{-2},$ while
$p\equiv\sqrt{p_{1}^{2}+p_{2}^{2}+p_{3}^{2}};$ the constant $\rho_{r}$ that
represents the radiation density has dimension of $\left(  length\right)
^{-4};$ finally $x_{c}^{0}\equiv-\frac{\sqrt{6}p}{\kappa\rho_{r}}$ is the time
of the crunch while the bang is set at zero time. This solution is plotted
parametrically in Fig. 2. The arrows on the curve show the evolution of the
fields $\left(  \phi_{\gamma}\left(  x^{0}\right)  ,s_{\gamma}\left(
x^{0}\right)  \right)  $ as conformal time $x^{0}$ increases from negative
values to positive values as the system goes through the crunch singularity at
$x^{0}=x_{c}^{0}$ and the bang singularity at $x^{0}=0.$%

\begin{figure}
[ptb]
\begin{center}
\includegraphics[
height=2.1577in,
width=2.1577in
]%
{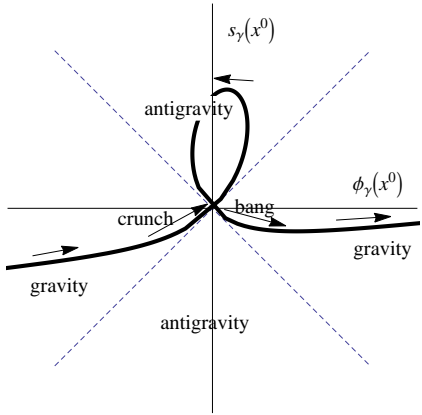}%
\caption{Geodesically complete spacetime requires the antigravity loop. The
solution for $\left(  \phi_{\gamma},s_{\gamma},g_{\mu\nu}^{\gamma}\right)  $
is continuous at the crunch $x^{0}=x_{c}^{0}<0$ and the bang $x^{0}=0.$ The
duration of the antigravity loop is $\left\vert x_{c}^{0}\right\vert
=\frac{\sqrt{6}p}{\kappa\rho_{r}}.$ Geodesics in this spacetime go smoothly
through both the crunch and the bang \cite{BST-sailing}. So information sails
through singularities in this spacetime. }%
\label{Fig2}%
\end{center}
\end{figure}
At the instants of crunch or bang the scale factor $a_{E}\left(  x^{0}\right)
$ in the \textit{Einstein gauge}, which in $d=4$ is given by
\cite{BCST1-antigravity}, $a_{E}^{2}=\frac{\kappa^{2}}{6}\left\vert
\phi_{\gamma}^{2}-s_{\gamma}^{2}\right\vert =\frac{\kappa^{2}}{6}\rho
_{r}\left\vert x^{0}(x^{0}-x_{c}^{0})\right\vert ,$ vanishes, and the
corresponding scalar curvature in the Einstein frame, $R\left(  g_{E}\right)
,$ blows up. However in the $\gamma$-gauge, since $a_{\gamma}=1,$ the behavior
is much milder since $a_{\gamma}$ is a constant, which is how we were able to
obtain our geodesically complete analytic solutions in the $\gamma$-gauge. If
the anisotropy coefficients $p_{1,2}$ are non-zero, then the Weyl invariant
Weyl curvature tensor, $C_{~\nu\lambda\sigma}^{\mu}\left(  g\right)  ,$ is
singular in all gauges, including in the $\gamma$-gauge. However, this fact
did not prevent us from obtaining \cite{BCST1-antigravity} our geodesically
complete solutions as given above explicitly, nor did it stop the geodesics in
this geometry from sailing through the cosmological singularities in the
presence of anisotropy \cite{BST-sailing}, thus showing that information does
go through singularities in our cosmological geometry in Eqs.(\ref{metric}%
-\ref{phi-s}). The underlying reason and technique for our ability to complete
the geometry and geodesics across these singularities is the identification of
a sufficient number of conserved quantities in our equations that allows us to
match continuously all observables, finite or infinite, across the
singularities \cite{BCST1-antigravity}\cite{BST-sailing}. The completion of
all geometrical features on both sides of singularities is evident since all
observables, including those gauge invariants that blow up such as
$C_{~\nu\lambda\sigma}^{\mu}\left(  g^{\gamma}\right)  ,$ are constructed from
our solutions for $\alpha_{1},\alpha_{2},\phi_{\gamma},s_{\gamma}$ which are
continuous through the singularities.

The solution displayed in Fig. 2 is generic, exact, and includes all possible
initial conditions (isotropic limit as well), hence there are no other
solutions in our setting (\ref{actionWinv2}) under the conditions stated above
(in particular, $V=0$). However, since $V$ cannot be neglected away from the
singularities this solution is useful mainly to study the neighborhood of the
singularities analytically, which is in fact our focus, and of course
precisely where string theory would play a role. So this geodesically complete
geometry is of great interest to string theory.

When $V\neq0,$ the complete set of solutions are known analytically for
certain potentials $V$ when anisotropy is neglected \cite{BCST2-solutions}.
The analytic expressions and their plots \cite{BCST2-solutions} are sufficient
to infer the solution for more general potentials $V.$ Combining the attractor
behavior near the singularities which is induced by anisotropy (Fig. 2)
together with the known behavior away from singularities when anisotropy is
negligible, the general generic behavior is pretty well understood as follows:
When $V$ is included, the generic solution is such that the trajectory shown
in Fig. 2 continues to move to large values of $\phi_{\gamma}$ in the gravity
region, while $s_{\gamma}$ oscillates at much smaller amplitudes (see
illustrations in \cite{BCT-cyclic},\cite{IB-cyclic},\cite{IB-BibBang}%
,\cite{BST-HiggsCosmo}). So, when the trajectory in the $\left(  \phi_{\gamma
},s_{\gamma}\right)  $ plane shown in Fig. 2 is continued to larger times
beyond the singularities, the curve turns around at large values of
$\phi_{\gamma}$ and comes back for another crunch at the origin of the
$\left(  \phi_{\gamma},s_{\gamma}\right)  $ plane, passes through another
antigravity loop and after another bang continues to the opposite gravity
region, doing this again and again an infinite number of times as conformal
time keeps progressing from minus infinity to plus infinity.

The Weyl invariant system in Eq.(\ref{actionWinv2}), coupled to ordinary
matter (here symbolized by the radiation parameter $\rho_{r}$), is then like a
perpetual motion machine, which absorbs energy from the gravitational field
and creates additional entropy during antigravity in each cycle
\cite{BST-HiggsCosmo}, thus yielding generically a cyclic universe, whose
global scale in the Einstein frame, $a_{E}\left(  x^{0}\right)  ,$ grows
progressively in each cycle due to entropy production, without having an
initial cycle in the multicycles in the far past or an end of multicycles in
the far future \cite{BST-HiggsCosmo}.

There are of course quantum corrections to this picture. Unfortunately, these
will remain obscure until quantum gravity is under better control. In the
interim, some new directions of research emerge from interesting new questions
that arise about the quantum physics during the antigravity period and close
to singularities. There are signals of instability because the kinetic energy
term for the gravitons is negative during antigravity. It seems gravitons
would be emitted copiously to transit to a lower energy state in response to
this instability. However, to maintain the energy constraints due to general
coordinate invariance ordinary matter must also be emitted simultaneously.
This is good for entropy production from one cycle to the next as discussed in
\cite{BST-HiggsCosmo}. Note that a measure of the amount of time spent in the
antigravity regime is $\left\vert x_{c}^{0}\right\vert =\frac{\sqrt{6}%
p}{\kappa\rho_{r}}$ as seen from our analytic solution (\ref{metric}%
-\ref{phi-s}) and Fig. 2 above. If radiation $\rho_{r}$ (any relativistic
matter) increases by entropy production during antigravity as mentioned above,
then the antigravity period $\left\vert x_{c}^{0}\right\vert $ gets shorter.
This is a signal that there seems to exist a built-in dynamical recovery
process to get from antigravity back to gravity. We hope that the string
version of our formulation below can shed light on the microscopic stringy
details of this mechanism.

The low energy theory has produced a wealth of signals for previously unknown
interesting phenomena. The general generic features we have noted in
\cite{BST-ConfCosm}-\cite{BST-sailing} are likely to survive in the context of
quantum gravity when we understand it better. To counter comments such as
those in \cite{Kallosh}\cite{Linde}, as we did in \cite{BST-sailing}, it may
be worth remembering that in string theory singularities tend to get smoother
or resolved, so we should expect less singular behavior as compared to the low
energy theory as stringy features are taken into account. It is not advisable
to try to include in our analysis stringy features in the form of higher
derivative corrections in the effective action. This is because those
corrections to the action are computed under the assumption of low energy away
from singularities, so they are the wrong tool to analyze phenomena near singularities.

Rather, we need to tackle string theory directly and ask again in that context
the types of questions we have been able to analyze and partially resolve in
the low energy theory. However, the starting point of string theory does not
seem to be well suited for our type of analysis since it begins with a
dimensionful constant, namely the string tension. For this reason we seek a
more general starting point, one in which the string tension is replaced by a
background field just like the gravitational constant which was replaced by a
field in the low energy theory. The construction of such a string theory is
the topic of the next section.

\section{Weyl Invariance and Dynamical Tension in String Theory}

The goal in this section is to develop a Weyl invariant formalism for strings
propagating in backgrounds, so that the string tension $\left(  2\pi
\alpha^{\prime}\right)  ^{-1}$ - equivalently the gravitational constant -
emerges from the gauge fixing of a background field in a new string theory
that has a local scaling gauge symmetry acting on \textit{target space} fields.

Taking the clues developed in the previous section we propose the following
string action with a dynamical string tension and target-space Weyl symmetry.
We simply insert in (\ref{Sstring}) our expressions for the Weyl covariant
substitutes for $2\alpha^{\prime}$ and $\Phi$ in terms of $\left(
\phi,s\right)  $ as given in Eq.(\ref{T}) and Eq.(\ref{s-gauge}). The result
is
\begin{equation}
S=\int d^{2}\sigma\left[
\begin{array}
[c]{c}%
-\frac{1}{2}T\left(  \phi\left(  X\right)  ,s\left(  X\right)  \right)
\left(  \sqrt{-h}h^{ab}g_{\mu\nu}\left(  X\right)  +\varepsilon^{ab}b_{\mu\nu
}\left(  X\right)  \right)  \partial_{a}X^{\mu}\partial_{b}X^{\nu}\\
+\frac{\sqrt{d-1}}{16\pi}\sqrt{-h}R^{\left(  2\right)  }\left(  h\right)
\ln\left(  \frac{\phi\left(  X\right)  +s\left(  X\right)  }{\phi\left(
X\right)  -s\left(  X\right)  }\right)  ^{2}+O\left(  \frac{1}{T}\right)
\end{array}
\right]  \label{SstringWeyl}%
\end{equation}
where $X^{\mu}\left(  \tau,\sigma\right)  $ is the string coordinate. If we
insert the s-gauge expressions for $\left(  \phi_{s},s_{s},g_{\mu\nu}%
^{s},b_{\mu\nu}^{s}\right)  $ given in Eq.(\ref{s-gauge}) for the $\pm$
gravity patches, this action reduces to the standard string action in
Eq.(\ref{Sstring}). For this more general action to be consistent with the
\textit{worldsheet conformal symmetry} up to order $\left(  1/T\right)  ,$ the
target space Weyl covariant background fields $\left(  \phi,s,g_{\mu\nu
},b_{\mu\nu}\right)  $ must be classical solutions of the equations of motion
that are derived from the target space Weyl invariant effective action
(\ref{actionWinv2}). The relevant solutions were described in the previous section.

The appearance of the logarithm in the action (\ref{SstringWeyl}) might be a
cause for concern, since the Lagrangian density appears non-analytic. However,
since the Euler density $\sqrt{-h}R^{\left(  2\right)  }$ is a total
derivative, one can integrate by parts, so that the resulting Lagrangian
density, involving only first order derivatives, has no branch cuts and is
single-valued on field space.

This string action contains no dimensionful constants. It is Weyl invariant
when the target-space background fields $\left(  \phi,s,g_{\mu\nu},b_{\mu\nu
}\right)  $ are rescaled locally according to the rules in
Eq.(\ref{WeylTransf}). This is easily seen by noting that the combinations of
fields
\begin{equation}
\hat{g}_{\mu\nu}\equiv T\left(  \phi,s\right)  g_{\mu\nu},\;\hat{b}_{\mu\nu
}\equiv T\left(  \phi,s\right)  b_{\mu\nu},\;\frac{s}{\phi}
\label{weylInvariantBackgr}%
\end{equation}
that appear in the basic string action (\ref{SstringWeyl}) are Weyl invariant.

We will see below that they also obey an additional symmetry which we call
\textit{flip symmetry}, that allows us to extend $T\left(  \phi,s\right)  ,$
from the $\pm$ gravity patches where it was initially defined in (\ref{T}), to
the geodesically complete entire $\left(  \phi,s\right)  $ plane.

Therefore it is possible to Weyl gauge fix the action in various ways (e.g.
$s,E,c,\gamma,f$ gauges of the previous section) to investigate the physics of
string theory in the corresponding spacetimes that look like different
backgrounds from one another. Any results obtained in one Weyl gauge are
easily transformed to another gauge, just like duality transformations (which
are now realized as Weyl transformations as explained at the end of the last section).

This expression for the string action is so far uniquely determined in the
gravity patches $\left(  \phi^{2}-s^{2}\right)  \geq0$ where the dynamical
tension $T\left(  \phi,s\right)  $ is positive. In this sense, with
Eq.(\ref{SstringWeyl}), we have already achieved a non-trivial stage for the
\textit{target-space Weyl invariant} definition of string theory. The
remaining questions involve the sign of the dynamical tension $T\left(
\phi,s\right)  $ in the antigravity patches so that the theory is sensible
according to sacred principles, such as unitarity, since a negative sign of
$T$ in (\ref{SstringWeyl}) may imply negative norm states. We will show below
that unitarity is not a problem, but there are physics questions to consider.
We emphasize that the issue is subtle because the antigravity patches occur
only for a finite amount of time $\left\vert x_{c}^{0}\right\vert $ in our
cosmological solutions and during that period those space-time patches are
separated by cosmological singularities from our own observable gravity patch,
as seen in Fig. 2.

In considering the sign of $T\left(  \phi,s\right)  $ in the antigravity
patches, there is an additional clue in the low energy Weyl invariant action
(\ref{actionWinv2}) to take into account. We observe that all the
\textit{kinetic} terms of the action in (\ref{actionWinv2}) flip sign if
$\phi$ and $s$ are interchanged with each other. Independently, if the metric
$g_{\mu\nu}$ flips signature, $g_{\mu\nu}\rightarrow-g_{\mu\nu},$ then noting
$R\left(  -g\right)  =-R\left(  g\right)  ,$ and that $\partial\phi
\cdot\partial\phi,$ $\partial s\cdot\partial s,$ $H^{2}$ all contain an odd
power of $g^{\mu\nu},$ while $\left(  -\det g\right)  $ is invariant for even
$d,$ we see that all kinetic terms in (\ref{actionWinv2}) flip sign under the
signature flip. Therefore, under the simultaneous flip of signature and
interchange of $\left(  \phi,s\right)  $
\begin{equation}
\text{flip symmetry:\ }\left(  \phi\leftrightarrow s\right)  \text{ and
}g_{\mu\nu}\rightarrow-g_{\mu\nu}, \label{flip}%
\end{equation}
all kinetic terms of the low energy action (\ref{actionWinv2}) are invariant
for even $d$. It is possible to include a flip of sign of $b_{\mu\nu
}\rightarrow-b_{\mu\nu},$ as an additional part of the flip transformations in
(\ref{flip}), since the flip of $b_{\mu\nu}$ is also a symmetry all by itself
in the low energy action. Observe also that $H_{\mu\nu\lambda}$ as given in
(\ref{HT}) is automatically flip symmetric for the $T$ given in (\ref{T}). The
remaining term in (\ref{actionWinv2}) is the potential energy $V\left(
\phi,s\right)  ,$ whose properties under the flip transformation is unclear at
this stage but should be determined by the underlying string theory.

Since the low energy action has a flip symmetry (at least in its kinetic
terms) we expect that the underlying string theory action (\ref{SstringWeyl}),
which should be compatible with all properties of (\ref{actionWinv2}), should
be symmetric under the flip transformation of all its \textit{background
fields}. Requiring this symmetry uniquely determines the properties of
$T\left(  \phi,s\right)  $ in all gravity/antigravity patches because the
combinations $T\left(  \phi,s\right)  g_{\mu\nu}$ and $T\left(  \phi,s\right)
b_{\mu\nu}$ that appear in (\ref{SstringWeyl}) must now be required to be flip
symmetric. This determines uniquely that $T$ must be antisymmetric under the
interchange of $\left(  \phi,s\right)  $
\begin{equation}
T\left(  \phi,s\right)  =-T\left(  s,\phi\right)  .
\end{equation}
Therefore the continuation of the expression $T\left(  \phi,s\right)  $ in
(\ref{T}) from the original gravity patch to all other gravity/antigravity
patches must obey this requirement. This can be done by writing $T_{string}%
\left(  \phi,s\right)  =\left(  \phi^{2}-s^{2}\right)  |\frac{T\left(
\phi,s\right)  }{\phi^{2}-s^{2}}|$ thus extending the $T\left(  \phi,s\right)
$ in (\ref{T}) to the entire $\left(  \phi,s\right)  $ plane. In what follows
we will assume that this has been done, but to avoid clutter, we will continue
to use the symbol $T\left(  \phi,s\right)  $ to mean the properly continued
$T_{string}\left(  \phi,s\right)  .$ This completes the definition of the Weyl
symmetric string action in (\ref{SstringWeyl}).

\subsection{String in Geodesically Complete Cosmological Background}

We will investigate the proposed action (\ref{SstringWeyl}) for the
geodesically complete cosmological background in Eqs.(\ref{metric}%
-\ref{phi-s}) that includes both gravity and antigravity patches as shown in
Fig. 2. This background is consistent with worldsheet conformal invariance
since it is the generic solution of the equations of motion of the effective
action (\ref{actionWinv2}). This background, which is an exact solution when
$V=0,$ and otherwise is the correct approximation near the singularity even
when $V\neq0,$ is suitable for our purpose of analyzing what happens near the
gravity/antigravity transitions. The worldsheet conformal symmetry has been
insured for any dimension $d$ (to lowest order in $\alpha^{\prime}$). Since
the background in Eqs.(\ref{metric}-\ref{phi-s}) was computed in four
dimensions$,$ we will specialize to $d=4$ without losing the (perturbatively
valid) worldsheet conformal symmetry.

We first compute the dynamical string tension (\ref{T}) for $d=4$ for this
background using Eqs.(\ref{T},\ref{phi+s},\ref{phi-s})
\begin{equation}
T_{4}\left(  X^{0}\right)  =\frac{1}{6}\left(  \phi_{\gamma}^{2}-s_{\gamma
}^{2}\right)  \left\vert \frac{\phi_{\gamma}+s_{\gamma}}{\phi_{\gamma
}-s_{\gamma}}\right\vert ^{\sqrt{3}}=\frac{1}{3}\rho_{r}X^{0}(X^{0}-x_{c}%
^{0})\left(  \frac{1}{2}\left\vert \frac{X^{0}}{l_{3}^{4}\rho_{r}(X^{0}%
-x_{c}^{0})}\right\vert ^{p_{3}/p}\right)  ^{\sqrt{3}} \label{T4}%
\end{equation}
Here $X^{0}\left(  \tau,\sigma\right)  $ is the string coordinate which is the
conformal time in the given background geometry. This $T_{4}\left(
X^{0}\right)  $ is flip antisymmetric as discussed in the previous section.

The metric $g_{\mu\nu}^{\gamma}$ was given in Eq.(\ref{metric}). But we
emphasize that only the Weyl invariant combination $\hat{g}_{\mu\nu}=T\left(
\phi_{\gamma},s_{\gamma}\right)  g_{\mu\nu}^{\gamma}$ enters in our action,
with%
\begin{equation}
d\hat{s}^{2}=\hat{g}_{\mu\nu}dx^{\mu}dx^{\nu}=T\left(  x^{0}\right)  \left(
\begin{array}
[c]{c}%
-\left(  dx^{0}\right)  ^{2}+e^{2\left(  \alpha_{1}+\sqrt{3}\alpha_{2}\right)
}\left(  dx^{1}\right)  ^{2}\\
+e^{2\left(  \alpha_{1}-\sqrt{3}\alpha_{2}\right)  }\left(  dx^{2}\right)
^{2}+e^{-4\alpha_{1}}\left(  dx^{3}\right)  ^{2}%
\end{array}
\right)  . \label{g-hat}%
\end{equation}
Explicitly, this purely time (string $X^{0}$) dependent cosmological
background for a string theory with dynamical tension is
\begin{equation}%
\begin{array}
[c]{l}%
\hat{g}_{00}\left(  X\right)  =-\frac{\rho_{r}X^{0}(X^{0}-x_{c}^{0})}%
{2^{\sqrt{3}}3}\left\vert \frac{X^{0}}{X^{0}-x_{c}^{0}}\right\vert
^{\frac{\sqrt{3}p_{3}}{p}}\left(  \rho_{r}l_{3}^{4}\right)  ^{-\frac{\sqrt
{3}p_{3}}{p}},\;\\
\hat{g}_{11}\left(  X\right)  =\frac{\rho_{r}X^{0}(X^{0}-x_{c}^{0})}%
{2^{\sqrt{3}}3}\left\vert \frac{X^{0}}{X^{0}-x_{c}^{0}}\right\vert
^{\frac{p_{1}}{p}+\frac{\sqrt{3}p_{2}}{p}+\frac{\sqrt{3}p_{3}}{p}}\left(
\rho_{r}l_{1}^{4}\right)  ^{-\frac{p_{1}}{p}}\left(  \rho_{r}l_{2}^{4}\right)
^{-\frac{\sqrt{3}p_{2}}{p}}\left(  \rho_{r}l_{3}^{4}\right)  ^{-\frac{\sqrt
{3}p_{3}}{p}},\\
\hat{g}_{22}\left(  X\right)  =\frac{\rho_{r}X^{0}(X^{0}-x_{c}^{0})}%
{2^{\sqrt{3}}3}\left\vert \frac{X^{0}}{X^{0}-x_{c}^{0}}\right\vert
^{\frac{p_{1}}{p}-\frac{\sqrt{3}p_{2}}{p}+\frac{\sqrt{3}p_{3}}{p}}\left(
\rho_{r}l_{1}^{4}\right)  ^{-\frac{p_{1}}{p}}\left(  \rho_{r}l_{2}^{4}\right)
^{+\frac{\sqrt{3}p_{2}}{p}}\left(  \rho_{r}l_{3}^{4}\right)  ^{-\frac{\sqrt
{3}p_{3}}{p}},\\
\hat{g}_{33}\left(  X\right)  =\frac{\rho_{r}X^{0}(X^{0}-x_{c}^{0})}%
{2^{\sqrt{3}}3}\left\vert \frac{X^{0}}{X^{0}-x_{c}^{0}}\right\vert
^{-\frac{2p_{1}}{p}+\frac{\sqrt{3}p_{3}}{p}}\left(  \rho_{r}l_{1}^{4}\right)
^{\frac{2p_{1}}{p}}\left(  \rho_{r}l_{3}^{4}\right)  ^{-\frac{\sqrt{3}p_{3}%
}{p}}.
\end{array}
\label{g-hat2}%
\end{equation}
where the parameters $\left(  p_{1},p_{2,}p_{3},p,l_{1},l_{2},l_{3},\rho
,x_{c}^{0}\right)  $ are defined after Eq.(\ref{phi-s}).

Note that in the low energy effective theory the original metric $g_{\mu\nu}$
in (\ref{actionWinv2}) does not change sign as the trajectory in Fig. 2
transits from a gravity region to an antigravity region at cosmological
singularities. Instead, the dynamical gravitational \textquotedblleft
constant\textquotedblright\ $\left(  \phi^{2}-s^{2}\right)  ,$ that multiplies
the curvature in the low energy action (\ref{actionWinv2}), changes sign at
those singularities. However, in view of the flip symmetry (\ref{flip}) of the
geodesically complete low energy theory (\ref{actionWinv2}), the sign flip of
the factor $\left(  \phi^{2}-s^{2}\right)  $ can also be viewed equivalently
as being a signature flip of the effective metric $\hat{g}_{\mu\nu}$ defined
above, since $R\left(  -\hat{g}\right)  =-R\left(  \hat{g}\right)  $. Indeed
the effective metric $\hat{g}_{\mu\nu}\left(  X^{0}\right)  $ in
(\ref{g-hat2}) changes overall signature precisely at the cosmological
singularities. The period of antigravity (or opposite signature of $\hat
{g}_{\mu\nu}$) and its temporary duration, $\left\vert x_{c}^{0}\right\vert
=\frac{\sqrt{6}p}{\kappa\rho_{r}},$ is determined by the physical parameters
of the background above.

Just as we could study particle geodesics in this background in
\cite{BST-sailing}, we can also study string geodesics in this geodesically
complete geometry by solving the equations for $X^{\mu}\left(  \tau
,\sigma\right)  $ that follow from our action (\ref{SstringWeyl}). This topic
is partially explored in the next section.

\subsection{Unitarity with a Temporary Flip of the Dynamical String Tension}

In this section we study a formalism for solving generally the classical
solutions and the quantization of the string theory described by our action
(\ref{SstringWeyl}), and its specialization to a cosmological background given
in Eq.(\ref{g-hat2}). To simplify this investigation we will drop the quantum
correction terms for worldsheet conformal symmetry in (\ref{SstringWeyl}) as
well as the $b_{\mu\nu}$ background field since these are not the crucial
terms for our questions involving the temporary sign flip of the dynamical
string tension. Hence we will simply concentrate on the purely classical
string action which is modified only by the dynamical string tension%
\begin{equation}
S_{cl}=-\frac{1}{2}\int d^{2}\sigma\sqrt{-h}h^{ab}\hat{g}_{\mu\nu}\left(
X\right)  \partial_{a}X^{\mu}\partial_{b}X^{\nu},\label{Sclass}%
\end{equation}
where the tension is absorbed into the definition of $\hat{g}_{\mu\nu}%
=Tg_{\mu\nu},$ allowing $\hat{g}_{\mu\nu}$ to have a temporary signature flip.
We are interested in obtaining the solutions for strings in such backgrounds,
and quantizing them to answer the questions on unitarity and begin to
interpret the physics when there is a signature or tension flip. Since such
questions are more general than the specific background in (\ref{g-hat2}), we
will consider a more general $\hat{g}_{\mu\nu}.$ Much of the necessary
canonical formalism was developed in Ref.~\cite{TPS}. It is useful to
introduce the momentum density defined by $P_{\mu}^{a}\equiv\partial
S_{cl}/\partial\left(  \partial_{a}X^{\mu}\right)  ,$
\begin{equation}
P_{\mu}^{a}=-\sqrt{-h}h^{ab}\hat{g}_{\mu\nu}\partial_{b}X^{\nu},\;\Rightarrow
\partial_{b}X^{\nu}=-\frac{h_{ba}}{\sqrt{-h}}\hat{g}^{\nu\mu}P_{\mu}^{a},
\end{equation}
where the canonical conjugate to $X^{\mu}$ corresponds to the momentum density
$P_{\mu}^{\tau},$ i.e. when $a=\tau.$ The classical string equations of motion
and constraints that follow from this action are
\begin{equation}%
\begin{array}
[c]{l}%
\delta X^{\mu}:0=\partial_{a}\left(  \sqrt{-h}h^{ab}\hat{g}_{\mu\nu}%
\partial_{b}X^{\nu}\right)  -\frac{\sqrt{-h}h^{ab}}{2}\partial_{a}X^{\lambda
}\partial_{b}X^{\rho}\partial_{X^{\mu}}\left(  \hat{g}_{\lambda\rho}\right)
\\
\delta h_{ab}:0=\;\hat{g}^{\mu\nu}P_{\mu}^{a}P_{\nu}^{b}-\frac{1}{2}h^{\alpha
b}h_{cd}\hat{g}^{\mu\nu}P_{\mu}^{c}P_{\nu}^{d}~.
\end{array}
\label{eoms}%
\end{equation}
To make progress we consider the worldsheet gauge $h_{\tau\sigma}=0$ in which
$h_{ab}$ is diagonal. There is remaining worldsheet gauge symmetry to choose
also the gauge $X^{0}=\tau$, which we will do later. Although a diagonal
$h_{ab}~$contains two functions, the combination $\sqrt{-h}h^{ab}$ contains
only one function since its determinant is $-1.$ Hence we parametrize $h_{ab}$
as follows%
\begin{equation}
\sqrt{-h}h^{ab}=\left(
\genfrac{}{}{0pt}{}{-h}{0}%
\genfrac{}{}{0pt}{}{0}{1/h}%
\right)  .\label{h and 1/h}%
\end{equation}
and insert it in all the equations listed in (\ref{eoms}). In what follows we
will assume that the metric is diagonal as is the case in (\ref{g-hat2})%
\begin{equation}
\hat{g}_{\mu\nu}=T\times diag\left(  -1,e_{1}^{2},e_{2}^{2},e_{3}^{2}%
,\cdots\right)  .
\end{equation}
Then,%

\begin{equation}%
\begin{array}
[c]{l}%
P_{\mu}^{\tau}=h\hat{g}_{\mu\nu}\partial_{\tau}X^{\nu},\;P_{\mu}^{\sigma
}=-h^{-1}\hat{g}_{\mu\nu}\partial_{\sigma}X^{\nu}\\
\mu=i:\;0=\partial_{\tau}\left(  -h\hat{g}_{ii}\partial_{\tau}X^{i}\right)
+\partial_{\sigma}\left(  h^{-1}\hat{g}_{ii}\partial_{\sigma}X^{i}\right)
,\text{ no sum on }i\\
\mu=0:\;0=\left(
\begin{array}
[c]{c}%
\partial_{\tau}\left(  -h\hat{g}_{00}\partial_{\tau}X^{0}\right)
+\partial_{\sigma}\left(  h^{-1}\hat{g}_{00}\partial_{\sigma}X^{0}\right)  \\
-\frac{1}{2}\left(  -h\partial_{\tau}X^{\lambda}\partial_{\tau}X^{\rho}%
+h^{-1}\partial_{\sigma}X^{\lambda}\partial_{\sigma}X^{\rho}\right)
\partial_{X^{0}}\hat{g}_{\lambda\rho}%
\end{array}
\right)  \\
0=\;\left(  P^{\tau}\pm hP^{\sigma}\right)  ^{2}\text{ or }0=\left(
h\partial_{\tau}X^{\nu}\pm\partial_{\sigma}X^{\nu}\right)  ^{2},\text{ }%
\end{array}
\label{eoms-X0gauge}%
\end{equation}
The last constraint equations are solved as%
\begin{equation}
h=\varepsilon\left(  X\right)  \sqrt{\frac{g_{\mu\nu}\partial_{\sigma}X^{\mu
}\partial_{\sigma}X^{\nu}}{-g_{\mu\nu}\partial_{\tau}X^{\mu}\partial_{\tau
}X^{\nu}}},\text{ and }g_{\mu\nu}\partial_{\tau}X^{\mu}\partial_{\sigma}%
X^{\nu}=0,\label{h-epsilon}%
\end{equation}
Note that in these expressions we wrote $g_{\mu\nu}$ rather than $\hat{g}%
_{\mu\nu}=Tg_{\mu\nu}$ because the tension factor of $T\left(  X\right)  $
drops out. Here $\varepsilon\left(  X\right)  $ is a sign function that
emerges from taking the square root. Assuming $h_{ab}$ has a fixed signature
we must take $\varepsilon\left(  X\right)  =1$ so that the signature of the
metric in (\ref{h and 1/h}) remains fixed forever\footnote{We make some
remarks here on the possibility that the worldsheet may also be allowed to
flip signature. Although we will not pursue this to the end in this paper, it
could be important to keep it in mind. The general solution of the $h_{ab}$
equation in (\ref{eoms}) is expressed in several forms
\begin{equation}
h^{ab}\left(  X\right)  =\hat{g}^{\mu\nu}\left(  X\right)  P_{\mu}^{a}P_{\nu
}^{b}\times\omega^{-1}\left(  X\right)  ,\text{ or }h_{ab}\left(  X\right)
=\hat{g}_{\mu\nu}\left(  X\right)  \partial_{a}X^{\mu}\partial_{b}X^{\nu
}\times\omega\left(  X\right)  .\label{h-factor}%
\end{equation}
The undetermined conformal factor $\omega\left(  X\right)  $ of the worldsheet
metric in (\ref{h-factor}) is immaterial except for its sign because its
absolute value always cancels in the worldsheet-Weyl invariant combination
$\sqrt{-h}h^{ab}$ that appears in all equations. However, in view of the fact
that the tension factor $T$ in the action (\ref{Sclass}) can flip sign
temporarily, we may entertain the possibility that the hitherto undetermined
conformal factor $\omega\left(  X\right)  $ in (\ref{h-factor}) could also
flip sign, thus inducing $h^{ab}$ to flip or not to flip signature, perhaps
simultaneously with $T.$ Since only the combination $T\left(  X\right)
\sqrt{-h}h^{ab}g_{\mu\nu}$ appears everywhere, the sign flip of $T$ could be
associated with $\hat{g}_{\mu\nu}$ as already mentioned above, but it is also
possible that in the solution of the theory as in (\ref{h-factor}) the
flipping signature of $\hat{g}^{\mu\nu}$ could be canceled by a flipping sign
of the factor $\omega\left(  X\right)  .$ Considering such solutions of the
theory amounts to associating the sign of $T\left(  X\right)  $ with $h_{ab}$
to combine them into an $\hat{h}_{ab}\equiv sign(T)h_{ab}$ that flips
signature at the gravity/antigravity boundaries, while leaving the signature
of $g_{\mu\nu}$ unchanged. But to be able to admit such solutions, the string
theory in Eq.(\ref{Sclass}) needs to be defined from the outset such that its
worldsheet is permitted to have such properties. The signature flip of the
effective $\hat{h}_{ab}$ amounts to the interchange of the role of
$\tau,\sigma$ as timelike/spacelike coordinates on the worldsheet as soon as
the string crosses the gravity/antigravity boundary. It is worth emphasizing
that this should not create any ghost problems since the worldsheet still has
only one time coordinate before or after the flip. If the string theory is
defined to include such properties, then the quantization of the theory and
the computation of amplitudes via the string path integral would need to admit
the possibility of such worldsheets. Since this possibility has not, as far as
we know, been considered before, we shall not pursue it further here.}.

Now, if we choose also the timelike gauge, $X^{0}=\tau,$ we obtain the
following form for the Hamiltonian density $\mathcal{H}\left(  \tau
,\sigma\right)  ,$ which is the negative of the canonical conjugate to $X^{0}$
as given in Eq.(\ref{eoms-X0gauge}),
\begin{equation}
X^{0}=\tau\;\Rightarrow\text{ }\mathcal{H}\left(  \tau,\sigma\right)
=-P_{0}^{\tau}=h\left(  \tau,\sigma\right)  T\left(  \tau\right)  ,
\end{equation}
where we insert the solution for $h$ in Eq.(\ref{h-epsilon}) to obtain the
Hamiltonian density (with $\varepsilon(X)$ set to 1$)$
\begin{equation}
\mathcal{H}\left(  \tau,\sigma\right)  =T\left(  \tau\right)  \left(
\frac{e_{j}^{2}\left(  \tau\right)  \left(  \partial_{\sigma}X^{j}\right)
^{2}}{1-e_{j}^{2}\left(  \tau\right)  \left(  \partial_{\tau}X^{j}\right)
^{2}}\right)  ^{1/2},
\end{equation}
Here $T\left(  X^{0}\right)  $ and $e_{i}^{2}\left(  X^{0}\right)  $ are
backgrounds, such as those in our suggested cosmological background in
(\ref{g-hat2}), evaluated at $X^{0}=\tau$.

This Hamiltonian may also be expressed in terms of the canonical conjugate to
$X^{i},$ which is $P_{i}^{\tau},$ instead of the velocity $\partial_{\tau
}X^{i}.$ Using
\begin{equation}
P_{i}^{\tau}=h\hat{g}_{ii}\partial_{\tau}X^{i}=T\times\left(  \frac{e_{j}%
^{2}\left(  \partial_{\sigma}X^{j}\right)  ^{2}}{1-e_{j}^{2}\left(
\partial_{\tau}X^{j}\right)  ^{2}}\right)  ^{1/2}e_{i}^{2}\partial_{\tau}%
X^{i},
\end{equation}
the Hamiltonian density in terms of canonical conjugates is obtained as
\begin{equation}
\mathcal{H}\left(  \tau,\sigma\right)  =T\left(  \tau\right)  \left(
e_{j}^{-2}\left(  \tau\right)  \left(  P_{j}^{\tau}\right)  ^{2}+\frac
{T^{2}\left(  \tau\right)  }{\pi^{2}}e_{j}^{2}\left(  \tau\right)  \left(
\partial_{\sigma}X^{j}\right)  ^{2}\right)  ^{1/2}.\label{H-density}%
\end{equation}
while the velocity is also given in terms of canonical variables
\begin{equation}
\partial_{\tau}X^{i}\left(  \tau,\sigma\right)  =\frac{T\left(  \tau\right)
P_{i}^{\tau}\left(  \tau,\sigma\right)  }{e_{i}^{2}\left(  \tau\right)
\mathcal{H}\left(  \tau,\sigma\right)  }=\frac{P_{i}^{\tau}}{e_{i}^{2}}\left(
e_{j}^{-2}\left(  \tau\right)  \left(  P_{j}^{\tau}\right)  ^{2}+T^{2}\left(
\tau\right)  e_{j}^{2}\left(  \tau\right)  \left(  \partial_{\sigma}%
X^{j}\right)  ^{2}\right)  ^{-1/2}.
\end{equation}

Note that in a time dependent curved cosmological background there is no
translation symmetry in the time coordinate $X^{0}.$ Therefore we do not
expect that the time translation generator $P_{0}^{\tau}$ would be conserved
in the gauge $X^{0}=\tau.$ Hence the Hamiltonian density or the total energy
of the system which is given by the Hamiltonian $H,$
\begin{equation}
H\left(  \tau\right)  =\int d\sigma\mathcal{H}\left(  \tau,\sigma\right)  ,
\label{energy}%
\end{equation}
will generally be a\ function of time $\tau$ in any cosmological background.

Having dealt with the gauge fixed $X^{0}$ and its conjugate Hamiltonian, the
remaining independent dynamical degrees of freedom $X^{i}\left(  \tau
,\sigma\right)  $ satisfy the following equations of motion and constraints
\begin{align}
0 &  =-\partial_{\tau}\left(  \mathcal{H}e_{i}^{2}\partial_{\tau}X^{i}\right)
+\partial_{\sigma}\left(  \frac{e_{i}^{2}}{\mathcal{H}}\partial_{\sigma}%
X^{i}\right)  ,\text{ no sum on }i\\
0 &  =\partial_{\sigma}X^{j}P_{j}^{\tau}\text{ or }0=e_{j}^{2}\partial_{\tau
}X^{j}\partial_{\sigma}X^{j}\text{ },\text{ summed over }j
\end{align}
The meaning of the constraint is the imposition of the $\sigma$%
-reparametrization gauge symmetry on physical states.

The quantization of this system is now evident. The equal time quantum rules
are
\begin{equation}
\left[  X^{i}\left(  \tau,\sigma\right)  ,P_{j}^{\tau}\left(  \tau
,\sigma^{\prime}\right)  \right]  =i\delta\left(  \sigma-\sigma^{\prime
}\right)  .
\end{equation}
On the Hilbert space created by these canonical operators we must impose the
physical state constraints which must be normal ordered at a fixed time $\tau$%
\begin{equation}
\left(  :\partial_{\sigma}X^{i}\left(  \tau,\sigma\right)  P_{i}^{\tau}\left(
\tau,\sigma\right)  :\right)  |phys\rangle=0.
\end{equation}

This Hilbert space is clearly unitary, since we have shown that a unitary
gauge exists in which only space-like canonical degrees of freedom are
involved in the quantization of our system. Space-like degrees of freedom can
never create negative norm states and the evolution of the system is generated
by a Hermitian Hamiltonian. So, \textit{there are no issues with unitarity}.

\section{Outlook}

In this section we address some open problems and point out some tools and new
directions to make further progress.

We argued that there are no problems with unitarity. However, there is the
unfamiliar feature, that the energy carried by any string configuration as
predicted by Eq.(\ref{energy}) will become negative once the string (along
with the rest of the universe) enters the antigravity patch. Taking our theory
as a cosmological model, the negative energy is supposed to happen in between
two cycles in an infinitely cyclic universe, so the question relates in
particular to what happens just before our current big bang. This is precisely
where we hope string theory could shed some light on quantum gravity.

Is the negative energy during the antigravity period a problem like an
instability, or just a new physics feature to investigate? How can we tell
from our perspective as observers in the gravity patches separated from
antigravity by cosmological singularities? One way to investigate this is to
consider the analog of a scattering process in which we imagine a string probe
that travels to the antigravity patch and returns to the gravity patch as an
altered string that brings information from the antigravity region. Based on a
set of solutions of this nature, we have already verified that the behavior of
string solutions as they pass through the crunch/bang singularities is regular
and is physically sensible. This analysis will appear in a forthcoming paper.

From the classical and quantum discussion above it is clear that the single
string system represented by the Weyl invariant action (\ref{SstringWeyl}),
with the cosmological configuration (\ref{g-hat2}) in mind, could not be
unstable during antigravity. This is because when the energy is negative, it
is negative for all classical and all quantum configurations of the single
string during the entire antigravity period for the whole universe. There are
no positive energy states of the single string during antigravity. Based on an
adiabatic approximation of energy conservation, it is not possible to make a
transition to a lower energy state because it is not possible to conserve
energy adiabatically with only negative energy states. Hence there can be no
transitions among the states of a single string that would indicate an
instability in the system described by (\ref{SstringWeyl}).

What about a multi-string system that includes string-string interactions? A
hint is provided by the flip symmetry of the low energy action. We have seen
already that all the kinetic terms flip sign when either $g_{\mu\nu}$ flips
signature or when $\left(  \phi,s\right)  $ are interchanged. This
transformation can be used to infer the properties of the theory in the
antigravity regime. If all terms of the effective action flip sign under the
$\left(  \phi,s\right)  $ interchange, then we deduce that all equations of
motion, and hence all physics, is the same during antigravity as compared to
gravity, and therefore nothing strange should be expected during antigravity.
If the potential $V\left(  \phi,s\right)  $ does not flip sign or has no
definite flip symmetry, then after multiplying the overall action by an
overall minus sign, the system would behave as if the potential is effectively
time dependent when one compares the antigravity/gravity periods. Like in any
time-dependent system, this system would undergo transitions. However this
does not imply that any sacred principles of physics would be violated. Our
considerations based on the low energy action and the flip symmetry of the
kinetic terms are reassuring that the physics is still familiar, but we need
better tools to understand the stringy details.

For a more powerful tool to investigate string-string interactions we may turn
to string field theory (SFT). This relates to our new action
(\ref{SstringWeyl}) since the corresponding SFT is constructed by using a BRST
operator derived from the stress tensor for the Weyl invariant action
(\ref{SstringWeyl}). Therefore SFT provides a conceptual framework in which
the relevant issues may be considered in a complete formalism that includes
all the interactions. Although it is hard to compute in SFT, recent
improvements of its formulation \cite{SFT-BR} with a new Moyal-type star
product that can be used for any curved space, makes it possible to use this
as a tool for more progress to answer our questions in quantum gravity. For
our purpose here the general structure of the open string field theory action
\cite{WittenSFT} is already a guide, as follows. In the so called
\textquotedblleft Siegel gauge\textquotedblright\ the BRST operator boils down
to the kinetic operator given by the Virasoro operator $L_{0},$ while the
basic interaction is cubic as in Chern-Simons theory. After some
manipulations, the open string field theory action can be gauge fixed to the
following form of non-commutative field theory
\begin{equation}
S=Tr\left[  \frac{1}{2}AL_{0}A+\frac{g}{3}A\star A\star A\right]  .
\end{equation}
where $L_{0}=\int d\sigma\left[  \text{Stress Tensor}\right]  ,$ is the zeroth
Virasoro operator. In the new formalism \cite{SFT-BR} the string field $A$ is
a function of \textit{half} of the string phase space $A\left(  X_{+}^{\mu
}\left(  \sigma\right)  ,P_{-\mu}^{\tau}\left(  \sigma\right)  \right)  ,$ the
trace $Tr$ is integration in the half phase space, while the star product
$\star$ is the new \textit{background independent} Moyal star product (which
represents string joining/splitting) in this \textit{half} phase
space\footnote{The string phase space is $X^{\mu}\left(  \sigma\right)
,P_{\mu}^{\tau}\left(  \sigma\right)  $ as defined above, and taken at fixed
$\tau$ (no gauge choice for $X^{0}$). \textit{Half} of the string phase space
is $\left(  X_{+}^{\mu}\left(  \sigma\right)  ,P_{-\mu}^{\tau}\left(
\sigma\right)  \right)  $ where the $\pm$ indicate parts of those string
degrees of freedom that are symmetric and antisymmetric relative to the
midpoint of the open string at $\sigma=\pi/2$ \cite{SFT-BR}.}. The point here
is that $L_{0},$ as a function of the stress tensor of our theory in
(\ref{SstringWeyl}) written in terms of canonical variables $\left(  X^{\mu
}\left(  \sigma\right)  ,P_{\mu}^{\tau}\left(  \sigma\right)  \right)  $, is
the only part of SFT that contains the new information about the flipping
dynamical tension $T\left(  X^{0}\left(  \sigma\right)  \right)  $ as a
function of $X^{0}\left(  \sigma\right)  .$ Hence, this formalism is an arena
that can be used to further investigate our questions including string-string
interactions both perturbatively in the coupling $g$, and non-perturbatively
including the search for the ground state as in any interacting field theory
in the presence of interactions. In this context a first impression is that,
cosmologically the negative energies will occur not only for single strings
but for all strings in the universe at the same time, and hence there should
be no instabilities, although there should be new physical features.

It would be very useful to construct examples of worldsheet-conformally-exact
string models whose string tension is Weyl-lifted to a background field as in
Eq.(\ref{SstringWeyl}). This would greatly improve the $\alpha^{\prime}$
expansion of Eq.(\ref{Sstring}) or the corresponding $1/T$ expansion in
(\ref{SstringWeyl}), to include all orders, and would also be essential in
practice for the SFT approach discussed above. A starting point to pursue this
idea could be the (gauged) Wess-Zumino-Witten models like those discussed in
\cite{IB-WZWs}-\cite{tseytlin}. Ideally we would like to find a conformally
exact cosmological string background whose behavior near the cosmological
singularities matches the behavior explicitly computed in Eqs.(\ref{alpha1}%
-\ref{phi-s}) and Fig. 2, and whose effective metric near the cosmological
singularities matches the form of Eq.(\ref{g-hat2}) after including the
dynamical tension.

Pursuing duality concepts is bound to be helpful to clarify the questions
mentioned above and to generalize our Weyl invariance approach to other
corners of M-theory. We mentioned earlier the connection of our target space
Weyl symmetry to dualities within the bosonic string theory in connection to
scale inversions \cite{Veneziano} and the more general cases involving
transformations among the $\left(  s,E,c,\gamma,f\right)  $-type gauge choices
(see also footnote (\ref{foot2T})). There is more to consider on the duality
path by pursuing the flip symmetry (\ref{flip}) further. A version of our flip
symmetry was considered in \cite{Duff1}\cite{Duff2} for a variety of
supersymmetric and heterotic strings in the context of low energy string
theory, but without our Weyl lifting ideas. We emphasize that without Weyl
lifting such a symmetry is not really realizable in the standard formulation
of string theory because, as stated in \cite{Duff1}\cite{Duff2}, it involves
the sign flip of the effective string coupling $e^{\Phi}$ to $-e^{\Phi}$. This
cannot be achieved by any reasonable transformation of the dilaton $\Phi$ as a
field that takes values only on the real line. However, with our Weyl lifting,
there is also a factor related to the dynamical tension $T$ that multiplies
$e^{\Phi}$ instead of the constant tension. It is the extra factor, not
$\Phi,$ that transforms under our flip symmetry, and this is realized in our
case by the interchange of $\left(  \phi,s\right)  $. \ We may consider the
work in \cite{Duff1}\cite{Duff2} as a possible guide toward the generalization
of our Weyl lifted string theory (\ref{SstringWeyl}) to a variety of dual
corners of M-theory, including supersymmetric and heterotic versions, and to
pursue duality concepts.

The tools and ideas mentioned above offer concrete approaches to further
exploration of string theory with target-space Weyl invariance and a dynamical
string tension.

\bigskip

We acknowledge conversations with Ed Witten on this topic. This research was
partially supported by the U.S. Department of Energy under grant number
DE-FG03-84ER40168 (IB) and under grant number DE-FG02-91ER40671 (PJS).
Research at Perimeter Institute is supported by the Government of Canada
through Industry Canada and by the Province of Ontario through the Ministry of
Research and Innovation.

\end{document}